\documentclass[12pt,a4paper]{article}
\usepackage{amsmath,amssymb,amsthm}
\usepackage[margin=1.0in]{geometry}
\usepackage{cite}
\usepackage{graphicx}
\usepackage{enumerate}
\usepackage{epsfig}
\usepackage{textcomp}
\usepackage{float}
\usepackage{bm}
\usepackage[displaymath, mathlines]{lineno}
\allowdisplaybreaks[0]
\floatstyle{plaintop}
\restylefloat{table}
\usepackage[colorlinks=true
,urlcolor=blue
,anchorcolor=blue
,citecolor=blue
,filecolor=blue
,linkcolor=blue
,menucolor=blue
,linktocpage=true
,pdfproducer=medialab
,pdfa=true
]{hyperref}
\numberwithin{equation}{section}

\usepackage{adjustbox}
\usepackage{lipsum}
\usepackage{pbox}

\usepackage{graphicx}
\usepackage{subcaption}

\usepackage[font=footnotesize]{caption}

\def\a{\alpha}

\def\m{\mu}

\def\L{\Lambda}

\def\be{\begin{equation}}
\def\ee{\end{equation}}
\def\bea{\begin{eqnarray}}
\def\eea{\end{eqnarray}}

\def\pa{\partial}

\def\lp{\left(}
\def\rp{\right)}
\def\ls{\left[}
\def\rs{\right]}
\def\nn{\nonumber}
\def\ie{{\it i.e., }}

\makeatletter
\renewcommand\section{\@startsection {section}{1}{\z@}%
	{-3.5ex \@plus -1ex \@minus -.2ex}
	{2.3ex \@plus.2ex}%
	{\normalfont\large\bfseries}}
\renewcommand\subsection{\@startsection{subsection}{2}{\z@}%
	{-3.25ex\@plus -1ex \@minus -.2ex}%
	{1.5ex \@plus .2ex}%
	{\normalfont\bfseries}}
\makeatother

\begin{document}

\begin{center}
\addtolength{\baselineskip}{.5mm}
\thispagestyle{empty}
\begin{flushright}
\end{flushright}

\vspace{20mm}

{\Large \bf Joule-Thomson expansion and heat engine efficiency of charged rotating black strings}
\\[15mm]
{Hamid R. Bakhtiarizadeh\footnote{h.bakhtiarizadeh@kgut.ac.ir}}
\\[5mm]
{\it Department of Nanotechnology, Graduate University of Advanced Technology,\\ Kerman, Iran}

\vspace{20mm}

{\bf  Abstract}
\end{center}

We perform the first study of the throttling process and heat engine efficiency of asymptotically Anti-de Sitter charged and rotating black strings in the extended phase space. For the throttling process, we calculate the Joule-Thomson coefficient and inversion temperature. We also depict the inversion and isenthalpic curves in the temperature-pressure plane, thereby identifying the corresponding cooling and heating regions. For the black string heat engine, we obtain analytical expressions for the efficiency of the Carnot and rectangular engine cycles and draw their diagrams in terms of some relevant thermodynamics variables.

\vfill
\newpage


\section{Introduction}\label{int}

Over the past fifty years, black hole thermodynamics has emerged as one of the most captivating areas in modern theoretical physics. It reveals a profound and surprising connection between thermodynamics, classical gravity, and quantum mechanics, offering a potential pathway toward a deeper understanding of quantum gravity. This field began with the concept of black hole entropy \cite{Bekenstein:1973ur}, followed by the formulation of the laws of black hole thermodynamics \cite{Bardeen:1973gs}, and culminated in the groundbreaking discovery of Hawking radiation \cite{Hawking:1975vcx}. These developments have shown that black holes are far more than simple mechanical entities. They are complex thermodynamic systems characterized by both temperature and entropy.

Black holes can display complex thermodynamic properties, particularly within Anti-de Sitter (AdS) spacetime characterized by a negative cosmological constant. Initial investigations into AdS black hole thermodynamics focused on the Hawking-Page phase transition, which describes a shift between a stable Schwarzschild-AdS black hole and thermal radiation in AdS space \cite{Hawking:1982dh}. Subsequent studies expanded this analysis to include charged AdS black holes, namely, Reissner-Nordström-AdS (RN-AdS) black holes \cite{Chamblin:1999tk,Chamblin:1999hg,Banerjee:2011au,Banerjee:2011raa}, revealing a striking resemblance between their phase behavior and that of van der Waals fluids.

In recent years, growing interest has emerged around the idea of treating the cosmological constant not as a fixed parameter of the AdS background, but as a dynamic thermodynamic pressure and the thermodynamic volume is defined as its conjugate \cite{Caldarelli:1999xj,Kastor:2009wy,Kastor:2010gq,Dolan:2010ha,Dolan:2011xt,Cvetic:2010jb}. This approach effectively introduces an additional dimension to the black hole thermodynamic phase space, giving rise to what is known as ``black hole thermodynamics in the extended phase space." In a seminal work by Kubiznak and Mann \cite{Kubiznak:2012wp}, this framework was systematically explored alongside the critical behavior of RN-AdS black holes. Within this context, the black hole mass is interpreted as enthalpy rather than internal energy, and the cosmological constant $ \L $ is redefined as a thermodynamic pressure. This framework enables a direct analogy between RN-AdS black holes and van der Waals fluids, particularly through similarities such as the small-large black hole phase transitions mirroring the gas-liquid phase transitions. These intriguing parallels sparked a surge of subsequent research. It was soon revealed that not only RN-AdS black holes but also a wide range of other black hole solutions exhibit van der Waals-like behavior. Furthermore, various thermodynamic aspects, such as thermodynamic geometry \cite{Zhang:2015ova}, compressibility \cite{Dolan:2011jm}, heat engine cycles \cite{Wei:2017vqs}, Maxwell’s equal area construction \cite{Wei:2014qwa}, Ehrenfest equations \cite{Mo:2013ela}, triple points \cite{Altamirano:2013uqa}, reentrant phase transitions \cite{Altamirano:2013ane}, and critical phenomena \cite{Niu:2011tb}, have been extensively studied, especially within the context of modified gravity theories. For comprehensive reviews on recent advancements, see Refs.\cite{Altamirano:2014tva,Kubiznak:2016qmn}.

Among the various lines of research, the throttling process, also known as the Joule-Thomson (JT) effect, has emerged as a particularly intriguing thermodynamic phenomenon and has garnered increasing attention. In Ref. \cite{Okcu:2016tgt}, Ökcü and Aydıner were the first to investigate this effect for RN-AdS black holes, deriving both inversion and isenthalpic curves. They extended their analysis to Kerr–AdS black holes in Ref. \cite{Okcu:2017qgo}, obtaining comparable results. Related investigations have also been conducted in a variety of contexts, including RN–AdS black holes within higher-dimensional spacetimes \cite{Mo:2018rgq}, as well as in alternative theories such as Lovelock gravity \cite{Mo:2018qkt}, Gauss-Bonnet gravity \cite{Lan:2018nnp}, and Rainbow gravity \cite{MahdavianYekta:2019dwf}. Collectively, these studies have further validated the foundational findings reported in Ref. \cite{Okcu:2016tgt}. The throttling process of the four-dimensional rotating and charged AdS black holes, known as Kerr-Newman-AdS (KN-AdS) black holes is also examined in \cite{Zhao:2018kpz}. 

In 2014, Johnson \cite{Johnson:2014yja} computed the efficiency of a black hole acting as a heat engine. Building upon the framework he introduced, subsequent studies extended the analysis to various other black hole types, including Gauss-Bonnet (GB) black holes \cite{Johnson:2015ekr}, Born-Infeld (BI)-AdS black holes \cite{Johnson:2015fva}, dilatonic BI black holes \cite{Chandrasekhar:2016lbd}, rotating black holes \cite{Hennigar:2017apu}, charged AdS black holes \cite{Liu:2017baz}, Kerr-AdS and dyonic black holes \cite{Sadeghi:2015bcm}, BTZ black holes \cite{Mo:2017nhw}, polytropic black holes \cite{Setare:2015yra}, higher-dimensional AdS black holes \cite{Belhaj:2015hha}, black holes in conformal gravity \cite{Xu:2017ahm}, massive gravity black holes \cite{Hendi:2017bys}, benchmarking black holes \cite{Chakraborty:2017weq}, accelerating AdS black holes \cite{Zhang:2018vqs}, black holes in gravity’s rainbow \cite{EslamPanah:2018ums}, charged accelerating AdS black holes \cite{Zhang:2018hms}, nonlinear charged AdS black holes \cite{Nam:2019zyk}, charged rotating and accelerating AdS black holes \cite{EslamPanah:2019szt} and four-dimensional Einstein-GB-AdS black holes \cite{EslamPanah:2020hoj}.

Although the throttling process and heat engine efficiency of various AdS black holes have been extensively explored, a significant case remains unexamined, the asymptotically AdS, cylindrically symmetric, charged and rotating black holes, known as black strings found by Lemos \cite{Lemos:1994xp,Lemos:1995cm}. In this work, we are going to provide a comprehensive and general analysis of the throttling behavior and heat engine efficiency of asymptotically AdS charged and rotating black strings within the extended phase space framework. The thermodynamic properties of solutions in conventional phase space are investigated in \cite{Dehghani:2002rr} and they are generalized to the extended phase space in \cite{Bakhtiarizadeh:2025uks}. 

The structure of this paper is as follows. In Sec. \ref{therm}, we briefly outline the thermodynamic properties of asymptotically AdS charged and rotating black strings within the extended phase space. Sec. \ref{jte} is devoted to the throttling process of black strings, where we analyze the JT coefficient, inversion temperature, inversion and isenthalpic curves. In Sec. \ref{hee} we treat this black string as a heat engine. In this section, we compute the efficiency of the black string heat engine, and then compare it to the rectangular and Carnot efficiency. Finally, we present our conclusions in Sec. \ref{conc}. Throughout this work, we adopt natural units, setting $ G_{N}=c=\hbar=k_{B}=1 $.

\section{Thermodynamics in extended phase space}\label{therm}

In this section we are going to briefly review the thermodynamic properties of charged rotating black strings in extended phase space obtained in \cite{Bakhtiarizadeh:2025uks}. In four-dimensional spacetime, the metric corresponding to cylindrical or toroidal horizons takes the form \cite{Lemos:1994xp,Lemos:1995cm} 
\bea\label{met}
ds^2=-f(r)\lp \Xi dt -a d\phi \rp ^2+\frac{1}{f(r)}dr^2+\frac{r^2}{\ell^4} \lp a dt -\Xi \ell^2 d\phi \rp ^2+\frac{r^2}{\ell^2}dz^2,
\eea
where
\be\label{sileq}
\Xi=\sqrt{1+\frac{a^2}{\ell^2}},
\ee
and
\be\label{metfunc}
f(r)=\frac{r^2}{\ell^2}-\frac{m}{r}+\frac{q^2}{r^2}.
\ee
Here, the integration constants $ m $ and $ q $ are proportional to the mass and charge of the black string, respectively. The constants $ a $ and $ \ell $, both having dimensions of length, can be interpreted as the rotation parameter and AdS radius, respectively. In what follows, we focus on solutions exhibiting the cylindrical symmetry. This symmetry implies that the spacetime admits a commutative two-dimensional Lie group $ G_2 $. The time and radial coordinates range as $ -\infty<t<\infty, 0\leq r<\infty $, respectively. The possible horizon topologies can be classified as follows: \begin{enumerate}[(i)] \item A flat torus $ T^2 $ with topology $ S^1\times S^1 $ (\ie $ G_2=U(1) \times U(1) $), characterized by the coordinate ranges $ 0\leq \phi<2\pi,0\leq z<2\pi \ell $, corresponding to a closed black string, \item A cylinder with topology $ \mathbb{R}\times S^1 $ (\ie $ G_2=\mathbb{R} \times U(1) $), with coordinate ranges $ 0\leq \phi<2\pi,-\infty<z<\infty $, representing a stationary black string, \item An infinite plane with topology $ \mathbb{R}^2 $ defined by the ranges $ -\infty<\phi<\infty,-\infty<z<\infty $, which is non-rotating and can be interpreted as a black hole.\end{enumerate}
In this work, we focus exclusively on the first two cases, (i) and (ii).

The relevant thermodynamic potentials per unit length of black string horizon are given by \cite{Dehghani:2002rr}
\bea\label{thermopot}
{\cal M}=\frac{1}{16 \pi \ell}\lp 3 \Xi^2-1\rp m,\qquad T=\frac{3 r_{+}^{4}-q^2 \ell^2}{4 \pi r_{+}^{3} \ell^2 \Xi},\qquad
{\cal S}=\frac{ r_{+}^{2}\Xi}{4\ell},\qquad
\Omega=\frac{a}{\Xi \ell^2},
\nn\\{\cal J}=\frac{3}{16 \pi \ell} \Xi a  m,\qquad\qquad\qquad \Phi=\frac{q}{r_+\Xi},\qquad\quad {\cal Q}=\frac{q \Xi}{4 \pi\ell},\qquad m=\frac{r_+^3}{\ell^2}+\frac{ q^2}{r_+},
\eea
They also can be derived from \cite{Bakhtiarizadeh:2021vdo,Bakhtiarizadeh:2021hjr,Bakhtiarizadeh:2023mhk} by setting the coupling constants to zero. Accordingly, we refer to them as the mass parameter and the charge parameter. $ T $ is the Hawking temperature of the black string,  $ \Omega $ the angular velocity, $ \Phi $ the electrostatic potential difference between infinity and the horizon. Here also, $ {\cal M} $ is the mass, $ {\cal S} $ the entropy, $ {\cal J} $ the angular momentum and $ {\cal Q} $ the electric charge, per unit length of black string horizon.

The inclusion of the cosmological constant, $ \L $, naturally leads to its identification as the thermodynamic pressure, 
\be\label{pressure}
P=-\frac{\L}{8\pi}=\frac{3}{8\pi\ell^2},
\ee
and the thermodynamic volume of the black string per unit horizon length, $ {\cal V} $, is proposed to be defined as the thermodynamic variable conjugate to $ P $. 

The expression for mass $ {\cal M} $ in terms of thermodynamic variables $ ({\cal S}, P, {\cal J}, {\cal Q}) $ is given by \cite{Bakhtiarizadeh:2025uks}
\bea\label{mass}
&&\!\!\!\!\!\!\!\!\!\!{\cal M} =\frac{1}{2^{3/4} \sqrt{3} \sqrt[4]{\pi } \sqrt[4]{P} \sqrt{\mathcal{S}} \left(32 \pi ^{3/2} \sqrt{2} \mathcal{J}^2 P^{3/2} \mathcal{S}+\varUpsilon \right)^{3/2}} \left[32 \pi ^{3/2} \sqrt{2} \mathcal{J}^2 P^{3/2} \mathcal{S} \varUpsilon +36864 P^4 \mathcal{S}^8\right.\nn\\&&\!\!\!\!\!\!\!\!\!\!\left.+55296 \pi  P^3 \mathcal{Q}^2 \mathcal{S}^6+32 \pi ^3 P \mathcal{S}^2 \left(64 \mathcal{J}^4 P^2+243 \mathcal{Q}^6\right)+31104 \pi ^2 P^2 \mathcal{Q}^4 \mathcal{S}^4+729 \pi ^4 \mathcal{Q}^8 \right],
\eea
where
\bea\label{varUpsilon}
&&\!\!\!\!\!\!\!\!\!\!\varUpsilon\equiv\left[110592 P^4 \mathcal{S}^8+165888 \pi  P^3 \mathcal{Q}^2 \mathcal{S}^6+32 \pi ^3 P \mathcal{S}^2 \left(64 \mathcal{J}^4 P^2+729 \mathcal{Q}^6\right)\right.\nn\\ &&\left.+93312 \pi ^2 P^2 \mathcal{Q}^4 \mathcal{S}^4+2187 \pi ^4 \mathcal{Q}^8\right]^{1/2}.
\eea
It is straightforward to demonstrate that in terms of thermodynamic variables $ ({\cal S}, P, \cal J, \cal Q) $, we have \cite{Bakhtiarizadeh:2025uks}
\bea\label{thermovol}
&&\!\!\!\!\!\!\!\!\!\!\mathcal {V}=\lp\frac{\pa {\cal {\cal M}}}{\pa P}\rp_{\cal S,\cal J,\cal Q} = \frac{\sqrt{32 \pi ^{3/2} \sqrt{2} \mathcal{J}^2 P^{3/2} \mathcal{S}+\varUpsilon }}{108\ 2^{3/4} \sqrt{3} \sqrt[4]{\pi } P^{5/4} \sqrt{\mathcal{S}} \left(8 P \mathcal{S}^2+3 \pi  \mathcal{Q}^2\right)^5}\times\nn\\&&\left[256 \pi ^{5/2} \sqrt{2} \mathcal{J}^2 P^{3/2} \mathcal{Q}^2 \mathcal{S} \varUpsilon +884736 P^5 \mathcal{S}^{10}+1216512 \pi  P^4 \mathcal{Q}^2 \mathcal{S}^8+580608 \pi ^2 P^3 \mathcal{Q}^4 \mathcal{S}^6\right.\nn\\&&\left.-8 \pi ^4 P \mathcal{Q}^2 \mathcal{S}^2 \left(2048 \mathcal{J}^4 P^2+729 \mathcal{Q}^6\right)+93312 \pi ^3 P^2 \mathcal{Q}^6 \mathcal{S}^4-2187 \pi ^5 \mathcal{Q}^{10}\right],
\eea
\bea\label{temp}
&&\!\!\!\!\!\!\!\!\!\! T =\lp\frac{\pa {\cal M}}{\pa {\cal S}}\rp_{P,\cal J,\cal Q}= -\frac{9 \sqrt{3} \left(\pi  \mathcal{Q}^2-8 P \mathcal{S}^2\right) \left(8 P \mathcal{S}^2+3 \pi  \mathcal{Q}^2\right)^3 \left(32 \pi ^{3/2} \sqrt{2} \mathcal{J}^2 P^{3/2} \mathcal{S} \varUpsilon +\varUpsilon ^2\right)}{2\ 2^{3/4} \sqrt[4]{\pi } \sqrt[4]{P} \mathcal{S}^{3/2} \varUpsilon  \left(32 \pi ^{3/2} \sqrt{2} \mathcal{J}^2 P^{3/2} \mathcal{S}+\varUpsilon \right)^{5/2}},\nn\\ 
\eea
\bea
\Omega=\lp\frac{\pa {\cal M}}{\pa {\cal J}}\rp_{{\cal S},P,\cal Q} = 
\frac{16\ 2^{3/4} \pi ^{5/4} \mathcal{J} P^{5/4} \sqrt{\mathcal{S}}}{\sqrt{3} \sqrt{32 \pi ^{3/2} \sqrt{2} \mathcal{J}^2 P^{3/2} \mathcal{S}+\varUpsilon }},
\eea
\bea
&&\!\!\!\!\!\!\!\!\!\!\Phi =\lp\frac{\pa {\cal M}}{\pa \cal Q}\rp_{{\cal S}, P,\cal J}= \frac {\sqrt[4]{2} \pi ^{3/4} \mathcal{Q} \sqrt{32 \pi ^{3/2} \sqrt{2} \mathcal{J}^2 P^{3/2} \mathcal{S}+\varUpsilon }}{27 \sqrt{3} \sqrt[4]{P} \sqrt{\mathcal{S}} \left(8 P \mathcal{S}^2+3 \pi  \mathcal{Q}^2\right)^5}\times  \nn\\&&\!\!\!\!\!\!\!\!\!\!\left[-64 \sqrt{2} \pi ^{3/2} \mathcal{J}^2 P^{3/2} \mathcal{S} \varUpsilon +110592 P^4 \mathcal{S}^8+165888 \pi  P^3 \mathcal{Q}^2 \mathcal{S}^6\right.\nn\\&&\left.+32 \pi ^3 P \mathcal{S}^2 \left(128 \mathcal{J}^4 P^2+729 \mathcal{Q}^6\right)+93312 \pi ^2 P^2 \mathcal{Q}^4 \mathcal{S}^4+2187 \pi ^4 \mathcal{Q}^8\right].
\eea 
By transforming back to the geometric variables $ (r_+, a ({\rm or}~\Xi), q, \ell) $, one finds the corresponding values for thermodynamic potentials obtained in Eq. (\ref{thermopot}) in terms of geometric variables. 

By incorporating pressure, the first law of black hole thermodynamics should be modified to \cite{Bakhtiarizadeh:2025uks}:
\be\label{exfirstlaw}
d{\cal M} = Td{\cal S} + \Omega d{\cal J} + \Phi d{\cal Q}+{\cal V}dP.
\ee
The extended first law presented above aligns with an integrated expression that connects the black hole mass to other thermodynamic quantities, known as the Smarr relation:
\be\label{exsmarr}
0 = T {\cal S} + \Omega {\cal J} -2 P {\cal V}.
\ee
In writting the above result we have used the Euler’s formula, which states for homogeneous functions
\be
f (x, y, \cdots , z) \to f (\a^p x, \a^q y, \cdots , \a^r z) = \a^s f (x, y, \cdots , z),
\ee
and the derivative with respect to $ \a $ implies
\be
s f (x, y, \cdots , z)=p\lp\frac{\pa f}{\pa x}\rp x+q\lp\frac{\pa f}{\pa y}\rp y+\cdots+r\lp\frac{\pa f}{\pa z}\rp z.
\ee 
Replacing $ f $ with $ {\cal M} $ yields the Smarr relation (\ref{exsmarr}) \cite{Kastor:2009wy,Caldarelli:1999xj}, since the respective scaling dimensions of $ \lp{\cal M},{\cal S},{\cal J},{\cal Q},P\rp $ are $ \lp L^0,L^1,L^1,L^0,L^{-2}\rp $, respectively. 

\section{Joule-Thomson expansion}\label{jte}

In this section we are going to investigate the JT expansion for asymptotically AdS charged and rotating black strings presented in the previous section. In conventional thermodynamics, a throttling process refers to an adiabatic expansion in which a non-ideal fluid (gas or liquid) is forced through a valve or porous plug due to a pressure difference. During this process, the fluid may undergo a temperature change as it expands from a high-pressure region to a low-pressure one. Although the process is irreversible, the enthalpy remains constant between the initial and final states. Therefore, the expansion can be treated as an isenthalpic process to analyze the resulting temperature change. This temperature variation is characterized by the JT coefficient,
\bea\label{JTC}
\m=\lp \frac{\pa T}{\pa P} \rp_H=\frac{1}{{\cal C}_P}\ls T \lp\frac{\pa {\cal V}}{\pa T} \rp_P -{\cal V} \rs.
\eea
Here, the subscript $ H $ denotes that the partial derivative is taken along an isenthalpic process, which will be denoted by $ {\cal M} $ in the extended phase space. $ {{\cal C}_P} $ is the heat capacity at constant pressure per unit horizon length \cite{Bakhtiarizadeh:2025uks},
\bea\label{hcacp}
{\cal C}_{P}\!\!\!\!\!&=&\!\!\!\!\!T\lp \frac{\pa T}{\pa {\cal S}} \rp_{P,{\cal J},{\cal Q}}^{-1}\nn\\\!\!\!\!\!&=&\!\!\!\!\!-2 \mathcal{S} \left(\pi  \mathcal{Q}^2-8 P \mathcal{S}^2\right) \left(8 P \mathcal{S}^2+3 \pi  \mathcal{Q}^2\right) \left[-288 \sqrt{2} \pi ^{7/2} \mathcal{J}^2 P^{3/2} \mathcal{Q}^4 \mathcal{S} \varUpsilon \right.\nn\\&&\!\!\!\!\!\left.+4608 \pi ^{5/2} \sqrt{2} \mathcal{J}^2 P^{5/2} \mathcal{Q}^2 \mathcal{S}^3 \varUpsilon -18432 \sqrt{2} \pi ^{3/2} \mathcal{J}^2 P^{7/2} \mathcal{S}^5 \varUpsilon +7077888 P^6 \mathcal{S}^{12}\right.\nn\\&&\!\!\!\!\!\left.+15925248 \pi  P^5 \mathcal{Q}^2 \mathcal{S}^{10}+14929920 \pi ^2 P^4 \mathcal{Q}^4 \mathcal{S}^8+144 \pi ^5 P \mathcal{Q}^4 \mathcal{S}^2 \left(128 \mathcal{J}^4 P^2+2187 \mathcal{Q}^6\right)\right.\nn\\&&\!\!\!\!\!\left.+192 \pi ^4 P^2 \mathcal{Q}^2 \mathcal{S}^4 \left(512 \mathcal{J}^4 P^2+10935 \mathcal{Q}^6\right)+2048 \pi ^3 P^3 \mathcal{S}^6 \left(64 \mathcal{J}^4 P^2+3645 \mathcal{Q}^6\right)\right.\nn\\&&\!\!\!\!\!\left.+19683 \pi ^6 \mathcal{Q}^{12}\right]\left[452984832 P^8 \mathcal{S}^{16}+1358954496 \pi  P^7 \mathcal{Q}^2 \mathcal{S}^{14}+1783627776 \pi ^2 P^6 \mathcal{Q}^4 \mathcal{S}^{12}\right.\nn\\&&\!\!\!\!\!\left.+6912 \pi ^6 P^2 \mathcal{Q}^6 \mathcal{S}^4 \left(1024 \mathcal{J}^4 P^2+5103 \mathcal{Q}^6\right)+262144 \pi ^3 P^5 \mathcal{S}^{10} \left(5103 \mathcal{Q}^6-2560 \mathcal{J}^4 P^2\right)\right.\nn\\&&\!\!\!\!\!\left.+172032 \pi ^4 P^4 \mathcal{Q}^2 \mathcal{S}^8 \left(2048 \mathcal{J}^4 P^2+3645 \mathcal{Q}^6\right)+110592 \pi ^5 P^3 \mathcal{Q}^4 \mathcal{S}^6 \left(1701 \mathcal{Q}^6-512 \mathcal{J}^4 P^2\right)\right.\nn\\&&\!\!\!\!\!\left.+3779136 \pi ^7 P \mathcal{Q}^{14} \mathcal{S}^2+177147 \pi ^8 \mathcal{Q}^{16}\right]^{-1}.
\eea
When charge is zero,
\bea
{\cal C}_{P}(\mathcal{Q}=0)=\frac{\mathcal{S} \left(\pi ^3 \mathcal{J}^4-9 \pi ^{3/2} \mathcal{J}^2 \sqrt{\pi ^3 \mathcal{J}^4+54 P \mathcal{S}^6}+54 P \mathcal{S}^6\right)}{27 P \mathcal{S}^6-40 \pi ^3 \mathcal{J}^4}.
\eea 
When angular momentum is zero,
\bea
{\cal C}_{P}(\mathcal{J}=0)=\mathcal{S} \left(2-\frac{8 \pi  \mathcal{Q}^2}{8 P \mathcal{S}^2+3 \pi  \mathcal{Q}^2}\right).
\eea
When both charge and angular momentum are zero, 
\bea
{\cal C}_{P}(\mathcal{Q}=0,\mathcal{J}=0)=2 \mathcal{S}. 
\eea
\begin{figure}
	\centering
	\includegraphics{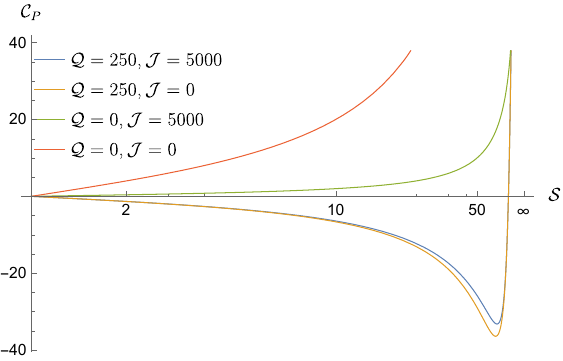}
	\caption{The profile of heat capacity at constant pressure as a function of entropy at constant pressure $ P=1 $.}\label{fig1}
\end{figure}
In Fig. \ref{fig1}, the profile of the specific heat per unit horizon length at constant pressure is illustrated as a function of entropy per unit horizon length. It can be seen, the specific heat for uncharged rotating black string is positive and therefore these solutions are thermodynamically stable, whereas the charged solutions are thermodynamically unstable except at large values of $ {\cal S} $. It also can be seen that there is a critical point for charged solutions, in contrast to the uncharged case, which occurs at the point of divergence of specific heat at constant pressure. This kind of divergence is a classical thermodynamic signal of a second-order phase transition for charged rotating black strings. These divergences typically occur at a critical point, analogous to the liquid-gas critical point in van der Waals fluids \cite{Bakhtiarizadeh:2025uks}.

Plugging Eqs. (\ref{hcacp}) and (\ref{thermovol}) into Eq. (\ref{JTC}), one finds the JT coefficient as
\bea
\m\!\!\!\!\!&=&\!\!\!\!\! - \ls 8817984 \pi ^{19/2} \sqrt{2} \mathcal{J}^2 P^{3/2} \mathcal{Q}^{16} \mathcal{S} \varUpsilon +172134236160 \pi ^{7/2} \sqrt{2} \mathcal{J}^2 P^{15/2} \mathcal{Q}^4 \mathcal{S}^{13} \varUpsilon \right.\\&&+289910292480 \pi ^{5/2} \sqrt{2} \mathcal{J}^2 P^{17/2} \mathcal{Q}^2 \mathcal{S}^{15} \varUpsilon +202937204736 \pi ^{3/2} \sqrt{2} \mathcal{J}^2 P^{19/2} \mathcal{S}^{17} \varUpsilon \nn\\&&-50096498540544 P^{12} \mathcal{S}^{24}-175337744891904 \pi  P^{11} \mathcal{Q}^2 \mathcal{S}^{22}\nn\\&&-270834195234816 \pi ^2 P^{10} \mathcal{Q}^4 \mathcal{S}^{20}-5668704 \pi ^{11} P \mathcal{Q}^{16} \mathcal{S}^2 \left(64 \mathcal{J}^4 P^2+81 \mathcal{Q}^6\right)\nn\\&&-268435456 \sqrt{2} \pi ^{9/2} \mathcal{J}^2 P^{13/2} \mathcal{S}^{11} \varUpsilon  \left(40 \mathcal{J}^4 P^2-243 \mathcal{Q}^6\right)\nn\\&&+49152 \pi ^{15/2} \sqrt{2} \mathcal{J}^2 P^{7/2} \mathcal{Q}^6 \mathcal{S}^5 \varUpsilon  \left(2560 \mathcal{J}^4 P^2+19683 \mathcal{Q}^6\right)\nn\\&&+147456 \pi ^{17/2} \sqrt{2} \mathcal{J}^2 P^{5/2} \mathcal{Q}^8 \mathcal{S}^3 \varUpsilon  \left(32 \mathcal{J}^4 P^2+729 \mathcal{Q}^6\right)\nn\\&&+524288 \pi ^{13/2} \sqrt{2} \mathcal{J}^2 P^{9/2} \mathcal{Q}^4 \mathcal{S}^7 \varUpsilon  \left(10935 \mathcal{Q}^6-1792 \mathcal{J}^4 P^2\right)\nn\\&&+6815744 \pi ^{11/2} \sqrt{2} \mathcal{J}^2 P^{11/2} \mathcal{Q}^2 \mathcal{S}^9 \varUpsilon  \left(1024 \mathcal{J}^4 P^2+3159 \mathcal{Q}^6\right)\nn\\&&-594316099584 \pi ^3 P^9 \mathcal{S}^{18} \left(64 \mathcal{J}^4 P^2+405 \mathcal{Q}^6\right)-43046721 \pi ^{12} \mathcal{Q}^{24}\nn\\&&-452984832 \pi ^4 P^8 \mathcal{Q}^2 \mathcal{S}^{16} \left(94208 \mathcal{J}^4 P^2+295245 \mathcal{Q}^6\right)\nn\\&&-452984832 \pi ^5 P^7 \mathcal{Q}^4 \mathcal{S}^{14} \left(29824 \mathcal{J}^4 P^2+102789 \mathcal{Q}^6\right)\nn\\&&-1048576 \pi ^6 P^6 \mathcal{S}^{12} \left(655360 \mathcal{J}^8 P^4+8680203 \mathcal{Q}^{12}\right)\nn\\&&-1152 \pi ^{10} P^2 \mathcal{Q}^8 \mathcal{S}^4 \left(262144 \mathcal{J}^8 P^4-2985984 \mathcal{J}^4 P^2 \mathcal{Q}^6-1594323 \mathcal{Q}^{12}\right)\nn\\&&-4096 \pi ^8 P^4 \mathcal{Q}^4 \mathcal{S}^8 \left(13631488 \mathcal{J}^8 P^4+14929920 \mathcal{J}^4 P^2 \mathcal{Q}^6-55801305 \mathcal{Q}^{12}\right)\nn\\&&+262144 \pi ^7 P^5 \mathcal{Q}^2 \mathcal{S}^{10} \left(1048576 \mathcal{J}^8 P^4+1135296 \mathcal{J}^4 P^2 \mathcal{Q}^6-1594323 \mathcal{Q}^{12}\right)\nn\\&&\left.+6144 \pi ^9 P^3 \mathcal{Q}^6 \mathcal{S}^6 \left(1048576 \mathcal{J}^8 P^4+559872 \mathcal{J}^4 P^2 \mathcal{Q}^6+7971615 \mathcal{Q}^{12}\right) \rs \nn\\&& \left\{ 4\ 2^{3/4} \sqrt{3} \sqrt[4]{\pi } P^{5/4} \mathcal{S}^{3/2} \varUpsilon  \left(8 P \mathcal{S}^2-\pi  \mathcal{Q}^2\right) \left(8 P \mathcal{S}^2+3 \pi  \mathcal{Q}^2\right)\times \right.\nn\\&&\sqrt{32 \pi ^{3/2} \sqrt{2} \mathcal{J}^2 P^{3/2} \mathcal{S}+\varUpsilon }  \ls-288 \sqrt{2} \pi ^{7/2} \mathcal{J}^2 P^{3/2} \mathcal{Q}^4 \mathcal{S} \varUpsilon +4608 \pi ^{5/2} \sqrt{2} \mathcal{J}^2 P^{5/2} \mathcal{Q}^2 \mathcal{S}^3 \varUpsilon \right.\nn\\&&-18432 \sqrt{2} \pi ^{3/2} \mathcal{J}^2 P^{7/2} \mathcal{S}^5 \varUpsilon +7077888 P^6 \mathcal{S}^{12}+15925248 \pi  P^5 \mathcal{Q}^2 \mathcal{S}^{10}\nn\\&&+14929920 \pi ^2 P^4 \mathcal{Q}^4 \mathcal{S}^8+144 \pi ^5 P \mathcal{Q}^4 \mathcal{S}^2 \left(128 \mathcal{J}^4 P^2+2187 \mathcal{Q}^6\right)+19683 \pi ^6 \mathcal{Q}^{12}\nn\\&&\left.\left.+192 \pi ^4 P^2 \mathcal{Q}^2 \mathcal{S}^4 \left(512 \mathcal{J}^4 P^2+10935 \mathcal{Q}^6\right)+2048 \pi ^3 P^3 \mathcal{S}^6 \left(64 \mathcal{J}^4 P^2+3645 \mathcal{Q}^6\right)\rs \right\}^{-1}.\nn
\eea
When the solution carries no electric charge,
\bea
\m(\mathcal{Q}=0)=\frac{\left(\sqrt{\pi ^3 \mathcal{J}^4+54 P \mathcal{S}^6}-6 \pi ^{3/2} \mathcal{J}^2\right) \sqrt{\pi ^{3/2} \mathcal{J}^2+\sqrt{\pi ^3 \mathcal{J}^4+54 P \mathcal{S}^6}}}{\sqrt{3} \sqrt[4]{\pi } \sqrt{P} \mathcal{S} \sqrt{\pi ^3 \mathcal{J}^4+54 P \mathcal{S}^6}}.
\eea
In the absence of rotation,
\bea
\m(\mathcal{J}=0)=-\frac{\sqrt[4]{\frac{3}{\pi }} \left(\pi  \mathcal{Q}^2-8 P \mathcal{S}^2\right)}{4\ 2^{3/4} P^{5/4} \mathcal{S}^{3/2}}.
\eea
For a non-rotating, uncharged solution,
\bea
\m(\mathcal{Q}=0,\mathcal{J}=0)=\frac{\sqrt[4]{\frac{6}{\pi }} \sqrt{\mathcal{S}}}{\sqrt[4]{P}}. 
\eea
\begin{figure}
	\centering
	\includegraphics{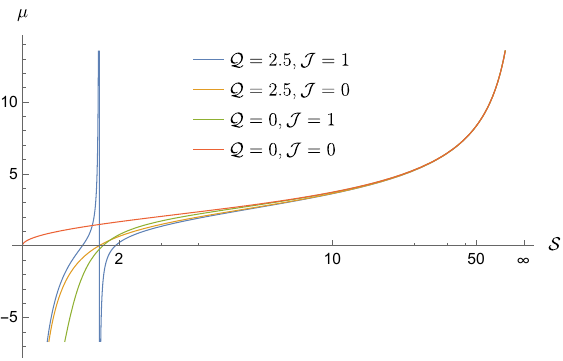}
	\caption{The profile of JT coefficient as a function of entropy at constant pressure $ P=1 $.}\label{fig2}
\end{figure}

In Fig. \ref{fig2}, the profile of JT coefficient as a function of entropy is illustrated. For a charged rotating configuration, it can be seen that $ \m $ monotonically increases and reaches $ 0 $ at $ {\cal S} = 1.21242 $ then it diverges at $ {\cal S} = 1.56664 $, where the Hawking temperature of the black string is zero. Then, it monotonically increases and reaches $ 0 $ at $ {\cal S} = 1.91858 $. Hence, there are only minimum inversion temperatures for the charged rotating black string, and the black strings always cool at large entropies.

Note that, the sign of $ \m $ divides the $ T-P $ plane into two distinct regions: a cooling region for which $ \m > 0 $ and a heating region at which $ \m < 0 $, since pressure consistently decreases during the throttling process. Clearly, when $ \m = 0 $, an inversion temperature $ T_i $ exists, defined as a solution to the equation:
\bea\label{invtempfor}
T_i={\cal V}\lp\frac{\pa T}{\pa {\cal V}}\rp_{P, {\cal J}, {\cal Q}}
\eea
and the relationship between $ T_i $ and $ P $ defines the inversion curve.

In the extended phase space, asymptotically AdS charged and rotating black strings exhibit thermodynamic behavior analogous to that of non-ideal fluids and can, therefore, undergo a throttling process. Naturally, there are no physical valves or porous plugs in the context of the universe; what is meant here is an adiabatic and isenthalpic process involving such black strings. This interpretation is physically well-motivated: the Hawking radiation emitted by a black string is typically negligible, justifying the assumption of adiabaticity. Furthermore, by neglecting radiation, the mass of the black string, which corresponds to enthalpy in the extended phase space, remains constant, rendering the process isenthalpic as well. Hence, both essential conditions for a throttling process are satisfied, just like what occurs for the KN-AdS black hole \cite{Zhao:2018kpz}. In what follows, we systematically examine the inversion temperature, inversion curve, and isenthalpic curves of asymptotically AdS charged and rotating black strings.

We can reexpress the inversion temperature in Eq. (\ref{invtempfor}) as
\bea\label{invtempfor2}
T_i={\cal V}\lp\frac{\pa T}{\pa {\cal V}}\rp_{P, {\cal J}, {\cal Q}}={\cal V}\frac{\lp \pa T/\pa {\cal S}\rp_{P, {\cal J}, {\cal Q}}}{\lp \pa {\cal V}/\pa {\cal S}\rp_{P, {\cal J}, {\cal Q}}}
\eea
Substituting Eqs. (\ref{thermovol}) and (\ref{temp}) into (\ref{invtempfor2}), we directly obtain the inversion temperature
\bea\label{invtemp}
T_i\!\!\!\!\!&=&\!\!\!\!\!-\ls-256 \sqrt{2} \pi ^{5/2} \mathcal{J}^2 P^{3/2} \mathcal{Q}^2 \mathcal{S} \varUpsilon -884736 P^5 \mathcal{S}^{10}-1216512 \pi  P^4 \mathcal{Q}^2 \mathcal{S}^8+2187 \pi ^5 \mathcal{Q}^{10}\right.\nn\\&&\left.+8 \pi ^4 P \mathcal{Q}^2 \mathcal{S}^2 \left(2048 \mathcal{J}^4 P^2+729 \mathcal{Q}^6\right)-93312 \pi ^3 P^2 \mathcal{Q}^6 \mathcal{S}^4-580608 \pi ^2 P^3 \mathcal{Q}^4 \mathcal{S}^6\rs\nn\\&& \ls1889568 \pi ^{15/2} \sqrt{2} \mathcal{J}^2 P^{3/2} \mathcal{Q}^{12} \mathcal{S}+16796160 \pi ^{13/2} \sqrt{2} \mathcal{J}^2 P^{5/2} \mathcal{Q}^{10} \mathcal{S}^3\right.\nn\\&&+94058496 \pi ^{11/2} \sqrt{2} \mathcal{J}^2 P^{7/2} \mathcal{Q}^8 \mathcal{S}^5+2491416576 \pi ^{3/2} \sqrt{2} \mathcal{J}^2 P^{15/2} \mathcal{S}^{13}\nn\\&&+2097152 \pi ^{5/2} \sqrt{2} \mathcal{J}^2 P^{13/2} \mathcal{S}^7 \left(40 \pi ^2 \mathcal{J}^4+1701 \mathcal{Q}^2 \mathcal{S}^4\right)\nn\\&&-393216 \sqrt{2} \pi ^{7/2} \mathcal{J}^2 P^{11/2} \mathcal{Q}^2 \mathcal{S}^5 \left(32 \pi ^2 \mathcal{J}^4-4941 \mathcal{Q}^2 \mathcal{S}^4\right)\nn\\&&+589824 \pi ^{9/2} \sqrt{2} \mathcal{J}^2 P^{9/2} \mathcal{Q}^4 \mathcal{S}^3 \left(4 \pi ^2 \mathcal{J}^4+891 \mathcal{Q}^2 \mathcal{S}^4\right)+19683 \pi ^6 \mathcal{Q}^{12} \varUpsilon\nn\\&&+7077888 P^6 \mathcal{S}^{12} \varUpsilon +65536 \pi  P^5 \mathcal{S}^6 \varUpsilon  \left(40 \pi ^2 \mathcal{J}^4+243 \mathcal{Q}^2 \mathcal{S}^4\right)\nn\\&&-12288 \pi ^2 P^4 \mathcal{Q}^2 \mathcal{S}^4 \varUpsilon  \left(32 \pi ^2 \mathcal{J}^4-1215 \mathcal{Q}^2 \mathcal{S}^4\right)+2099520 \pi ^4 P^2 \mathcal{Q}^8 \mathcal{S}^4 \varUpsilon\nn\\&&\left.+18432 \pi ^3 P^3 \mathcal{Q}^4 \mathcal{S}^2 \varUpsilon  \left(4 \pi ^2 \mathcal{J}^4+405 \mathcal{Q}^2 \mathcal{S}^4\right)+314928 \pi ^5 P \mathcal{Q}^{10} \mathcal{S}^2 \varUpsilon \rs\nn\\&&\left\{2\ 2^{3/4} \sqrt{3} \sqrt[4]{\pi } \sqrt[4]{P} \mathcal{S}^{3/2} \left(8 P \mathcal{S}^2+3 \pi  \mathcal{Q}^2\right) \left(32 \pi ^{3/2} \sqrt{2} \mathcal{J}^2 P^{3/2} \mathcal{S}+\varUpsilon \right)^{5/2} \right.\nn\\&& \ls 768 \pi ^{7/2} \sqrt{2} \mathcal{J}^2 P^{3/2} \mathcal{Q}^4 \mathcal{S} \varUpsilon -10240 \sqrt{2} \pi ^{5/2} \mathcal{J}^2 P^{5/2} \mathcal{Q}^2 \mathcal{S}^3 \varUpsilon +21233664 P^6 \mathcal{S}^{12}\right.\nn\\&&+40697856 \pi  P^5 \mathcal{Q}^2 \mathcal{S}^{10}+31518720 \pi ^2 P^4 \mathcal{Q}^4 \mathcal{S}^8+12441600 \pi ^3 P^3 \mathcal{Q}^6 \mathcal{S}^6+6561 \pi ^6 \mathcal{Q}^{12}\nn\\&&\left.\left.-48 \pi ^5 P \mathcal{Q}^4 \mathcal{S}^2 \left(1024 \mathcal{J}^4 P^2-5103 \mathcal{Q}^6\right)+320 \pi ^4 P^2 \mathcal{Q}^2 \mathcal{S}^4 \left(2048 \mathcal{J}^4 P^2+8019 \mathcal{Q}^6\right)\rs\right\}^{-1}.\nn\\
\eea
We begin by analyzing the minimum inversion temperature $ T_i^{min}({\cal J}, {\cal Q}) $, which corresponds to the inversion temperature at zero pressure. When $ P=0 $, Eq. (\ref{invtemp}) goes to
\be
T_i^{min}=0.
\ee
At this stage, however, $ T_i $ is still expressed as a function of $ P, {\cal J}, {\cal Q} $, and $ {\cal S} $. To express it explicitly in the form $ T_i = T_i(P, {\cal J}, {\cal Q}) $, we should extract the relation between $ {\cal S} $ and the parameters $ P, {\cal J} $, and $  {\cal Q} $. This relationship can be obtained through an additional mathematical manipulation. Specifically, we can rewrite Eq. (\ref{JTC}) as \cite{Zhao:2018kpz}:
\bea
0=T-{\cal V}\lp\frac{\pa T}{\pa {\cal V}}\rp_{P, {\cal J}, {\cal Q}}=T^2 \lp\frac{\pa \lp{\cal V}/T\rp}{\cal V}\rp_{P, {\cal J}, {\cal Q}}=T^2 \frac{\lp\pa\lp{\cal V}/T\rp/\pa {\cal S}\rp_{P, {\cal J}, {\cal Q}}}{\lp\pa \cal V/\pa {\cal S}\rp_{P, {\cal J}, {\cal Q}}}
\eea
Since a physical black hole always satisfies $ T\neq0 $ and $ \lp\pa \cal V/\pa {\cal S}\rp_{P, {\cal J}, {\cal Q}}\neq\infty $, it suffices to set the numerator equal to zero:
\bea\label{eq}
&&\!\!\!\!\!\varUpsilon \left(32 \pi ^{3/2} \sqrt{2} \mathcal{J}^2 P^{3/2} \mathcal{S}+\varUpsilon \right)^2 \left[839808 \pi ^{15/2} \sqrt{2} \mathcal{J}^2 P^{3/2} \mathcal{Q}^{12} \mathcal{S}\right.\\&&+8957952 \pi ^{13/2} \sqrt{2} \mathcal{J}^2 P^{5/2} \mathcal{Q}^{10} \mathcal{S}^3+53747712 \pi ^{11/2} \sqrt{2} \mathcal{J}^2 P^{7/2} \mathcal{Q}^8 \mathcal{S}^5\nn\\&&+905969664 \pi ^{3/2} \sqrt{2} \mathcal{J}^2 P^{15/2} \mathcal{S}^{13}+8388608 \pi ^{5/2} \sqrt{2} \mathcal{J}^2 P^{13/2} \mathcal{S}^7 \left(5 \pi ^2 \mathcal{J}^4+162 \mathcal{Q}^2 \mathcal{S}^4\right) \varUpsilon\nn\\&&-1572864 \sqrt{2} \pi ^{7/2} \mathcal{J}^2 P^{11/2} \mathcal{Q}^2 \mathcal{S}^5 \left(4 \pi ^2 \mathcal{J}^4-513 \mathcal{Q}^2 \mathcal{S}^4\right)-2187 \pi ^6 \mathcal{Q}^{12}\nn\\&&+1179648 \pi ^{9/2} \sqrt{2} \mathcal{J}^2 P^{9/2} \mathcal{Q}^4 \mathcal{S}^3 \left(\pi ^2 \mathcal{J}^4+216 \mathcal{Q}^2 \mathcal{S}^4\right)-7077888 P^6 \mathcal{S}^{12} \varUpsilon \nn\\&&+327680 \pi  P^5 \mathcal{S}^6 \varUpsilon  \left(4 \pi ^2 \mathcal{J}^4-27 \mathcal{Q}^2 \mathcal{S}^4\right)-12288 \pi ^2 P^4 \mathcal{Q}^2 \mathcal{S}^4 \varUpsilon  \left(16 \pi ^2 \mathcal{J}^4+279 \mathcal{Q}^2 \mathcal{S}^4\right)\nn\\&&\left.+18432 \pi ^3 P^3 \mathcal{Q}^4 \mathcal{S}^2 \varUpsilon  \left(2 \pi ^2 \mathcal{J}^4-9 \mathcal{Q}^2 \mathcal{S}^4\right)+139968 \pi ^4 P^2 \mathcal{Q}^8 \mathcal{S}^4 \varUpsilon +11664 \pi ^5 P \mathcal{Q}^{10} \mathcal{S}^2 \varUpsilon\right] =0.\nn
\eea
Unfortunately, this is an algebraic equation about $ \mathcal{S} $ of higher degree, so it is analytically unsolvable. Therefore, without loss of generality, we restrict ourselves to the uncharged case. In this case, the inversion temperature (\ref{invtemp}) takes the following form 
\bea\label{unchinvtemp}
T_i({\cal Q}=0)&=&\ls4 \sqrt{P} \left(10 \pi ^{9/2} \mathcal{J}^6+10 \pi ^3 \mathcal{J}^4 \sqrt{\pi ^3 \mathcal{J}^4+54 P \mathcal{S}^6}\right.\right.\nn\\&&\left.\left.+27 P \mathcal{S}^6 \sqrt{\pi ^3 \mathcal{J}^4+54 P \mathcal{S}^6}+297 \pi ^{3/2} \mathcal{J}^2 P \mathcal{S}^6\right)\rs\times\nn\\&&\ls3 \sqrt{3} \sqrt[4]{\pi } \mathcal{S} \left(\pi ^{3/2} \mathcal{J}^2+\sqrt{\pi ^3 \mathcal{J}^4+54 P \mathcal{S}^6}\right)^{5/2}\rs^{-1}.
\eea
The Eq. (\ref{eq}) also reduces to
\bea
27 P \left(1-\frac{4 \pi ^{3/2} \mathcal{J}^2}{\sqrt{\pi ^3 \mathcal{J}^4+54 P \mathcal{S}^6}}\right)-\frac{5 \pi ^3 \mathcal{J}^4+\frac{5 \pi ^{9/2} \mathcal{J}^6}{\sqrt{\pi ^3 \mathcal{J}^4+54 P \mathcal{S}^6}}}{\mathcal{S}^6}=0.
\eea
The above equation is analytically solvable. Among its six solutions, only one is positive and physically meaningful,
\be\label{unchent}
\mathcal{S}= \frac{\sqrt[6]{\frac{35}{2}} \sqrt{\pi } \mathcal{J}^{2/3}}{\sqrt{3} \sqrt[6]{P}}.
\ee
Substiuting Eq. (\ref{unchent}) into (\ref{unchinvtemp}), one can obtain the exact form of the inversion temperature for uncharged solutions
\be
T_i({\cal Q}=0)=\frac{2 \sqrt[6]{2} 5^{5/6} \sqrt[3]{\mathcal{J}} P^{2/3}}{7^{2/3}}.
\ee
\begin{figure}
	\centering
	\includegraphics{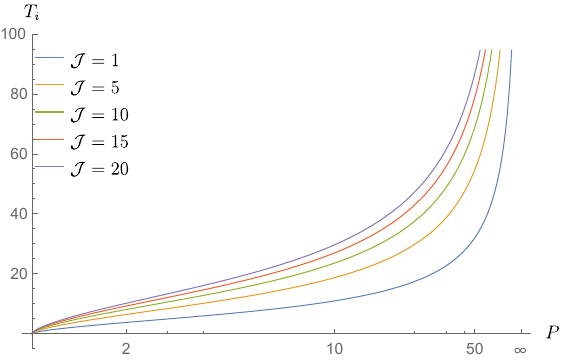}
	\caption{The inversion curves for $ {\cal Q}=0 $ and different values of $ {\cal J} $.}\label{fig3}
\end{figure}

In Fig. \ref{fig3}, the inversion curves for $ {\cal Q} = 0 $ and different values of $ {\cal J}  $ are plotted. In contrast to the van der Waals fluids, the inversion curves are not closed and there is only one inversion curve. A similar behaviour has been observed for RN-AdS \cite{Okcu:2016tgt}, Kerr-AdS \cite{Okcu:2017qgo} and KN-AdS \cite{Zhao:2018kpz} black holes. 
\begin{figure}
	\centering
	\begin{subfigure}[b]{.496\linewidth}
		\includegraphics[width=\linewidth]{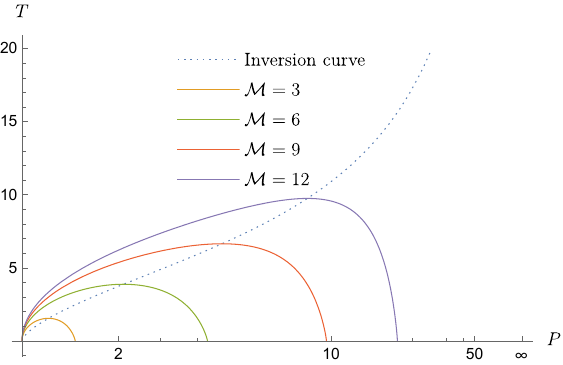}
		\setcounter{subfigure}{0}%
		\caption{$ {\cal J}=1 $}
	\end{subfigure}
	\begin{subfigure}[b]{.496\linewidth}
		\includegraphics[width=\linewidth]{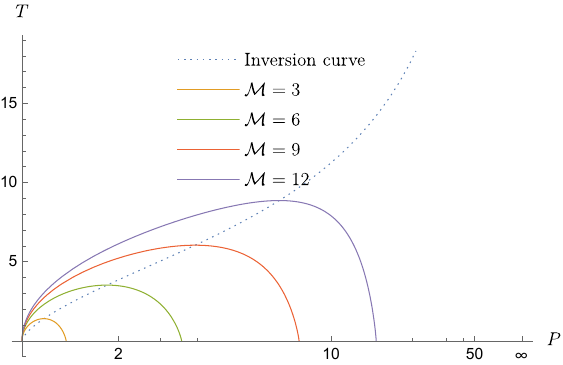}
		\caption{$ {\cal J}=1.1 $}
	\end{subfigure}
	\begin{subfigure}[b]{.496\linewidth}
		\includegraphics[width=\linewidth]{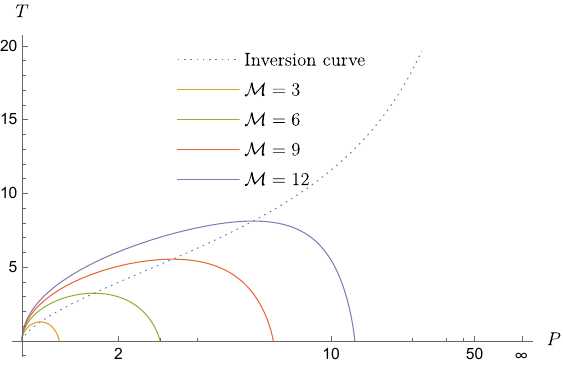}
		\caption{$ {\cal J}=1.2 $}
	\end{subfigure}
	\begin{subfigure}[b]{.496\linewidth}
		\includegraphics[width=\linewidth]{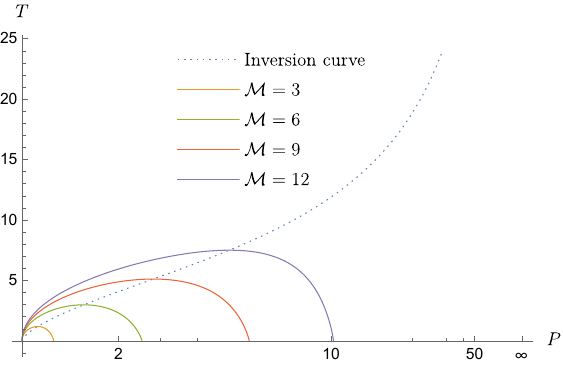}
		\caption{$ {\cal J}=1.3 $}
	\end{subfigure}
	\caption{The isenthalpic and inversion curves for uncharged solutions and different valuse of $ {\cal J} $ and $ {\cal M} $.}
	\label{fig4}
\end{figure}
In Fig. \ref{fig4}, the isenthalpic and inversion curves of uncharged solutions for various values of mass and angular momentum have been plotted in $ T-P $ plane. To draw the isenthalpic curves, we first derive the entropy from the mass formula (\ref{mass}), by setting $ {\cal Q}=0 $, as 
\bea\label{entinmass}
\mathcal{S}({\cal Q}=0)\!=\!\frac{\sqrt[6]{\pi }}{2\ 2^{2/3} \sqrt[3]{3} \sqrt{P}}\!\!\!\!\!\!\!\!\!\!&&\ls512 \pi ^2 \mathcal{J}^4 P^2-864 \pi  \mathcal{J}^2 P \mathcal{M}^2+243 \mathcal{M}^4+\frac{\mathcal{I}^3 \mathcal{M}}{3}\rs^{1/6},
\eea
where
\be
\mathcal{I}\equiv\sqrt{81 \mathcal{M}^2-192 \pi  \mathcal{J}^2 P}.
\ee
Plugging Eq. (\ref{entinmass}) into Eq. (\ref{temp}) and setting $ {\cal Q}=0 $ leads to the following value for the Hawking temperature of black string 
\bea
&&T({\cal Q}=0)=\nn\\&&\left\{6\ 2^{2/3} \sqrt[3]{3} \sqrt{P} \left(512 \pi ^2 \mathcal{J}^4 P^2-864 \pi  \mathcal{J}^2 P \mathcal{M}^2+\frac{\mathcal{I}^3 \mathcal{M}}{3}+243 \mathcal{M}^4\right)^{5/6}\right.\nn\\&& \left[2048 \pi ^2 \mathcal{J}^4 P^2+16 \pi  \mathcal{J}^2 P \left(\sqrt{\frac{2}{3}} \mathcal{I} \sqrt{3 \mathcal{M} (\mathcal{I}+9 \mathcal{M})-32 \pi  \mathcal{J}^2 P}-6 \mathcal{M} (2 \mathcal{I}+27 \mathcal{M})\right)\right.\nn\\&&\left.\left.+81 \mathcal{M}^3 (\mathcal{I}+9 \mathcal{M})\right]\right\}\left\{\sqrt[6]{\pi } \mathcal{I} \sqrt{3 \mathcal{M} (\mathcal{I}+9 \mathcal{M})-32 \pi  \mathcal{J}^2 P} \right.\nn\\&&\left.\left[32 \pi  \mathcal{J}^2 P+\sqrt{\frac{2}{3}} \mathcal{I} \sqrt{3 \mathcal{M} (\mathcal{I}+9 \mathcal{M})-32 \pi  \mathcal{J}^2 P}\right]^{5/2}\right\}^{-1}.
\eea
Thus we obtain the constant mass curves in $ T-P $ plane. As illustrated in Fig. \ref{fig4}, the area above the inversion curve corresponds to the cooling region, whereas the area below it represents the heating region. In fact, these regions can be identified by analyzing the sign of the slope of the isenthalpic curves: a positive slope indicates cooling, while a negative slope corresponds to heating. Importantly, no cooling or heating occurs on the inversion curve itself; it serves as the boundary separating the two regimes. 

\section{Heat engine efficiency}\label{hee}

When black strings are treated as thermodynamic systems within the extended phase space, it becomes natural to consider them as potential heat engines. In this context, the mechanical term $ Pd{\cal V} $ in the first law allows for the definition of mechanical work, and consequently, the efficiency of such engines. A heat engine operates along a closed cycle in the $ P-{\cal V} $ diagram and functions between two thermal reservoirs: a hot reservoir at temperature $ T_H $ and a cold reservoir at temperature $ T_C $. During the cycle, the engine absorbs a quantity of heat $ Q_H $ from the hot reservoir. A portion of this heat is converted into mechanical work $ W $, while the remaining heat $ Q_C $ is released to the cold reservoir. The efficiency of the heat engine is defined as:
\bea\label{effi}
\eta&\equiv&\frac{W}{Q_H}\nn\\&=&1-\frac{Q_C}{Q_H}.
\eea
The efficiency of a heat engine depends on the equation of state of the black hole and the specific thermodynamic paths that constitute the heat cycle in the $ P-{\cal V} $ diagram. 

\begin{figure}
	\centering
	\begin{subfigure}[b]{.37\linewidth}
		\includegraphics[width=\linewidth]{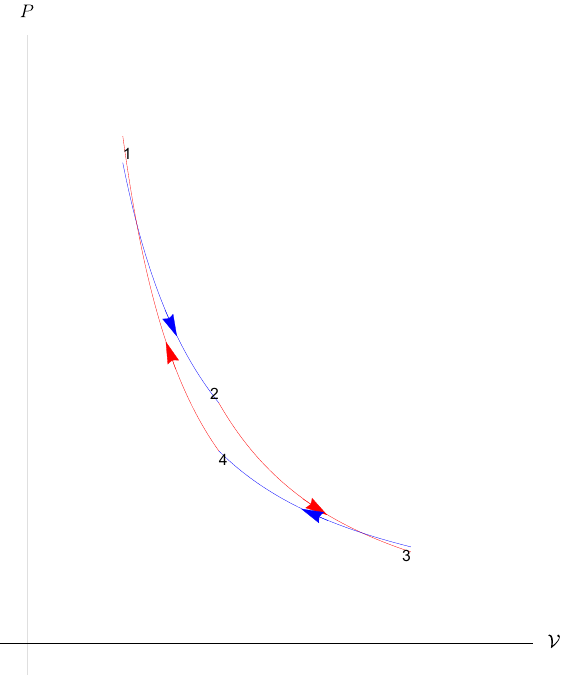}
		\setcounter{subfigure}{0}%
		\caption{Carnot engine}
	\end{subfigure}
	\begin{subfigure}[b]{.3\linewidth}
		\includegraphics[width=\linewidth]{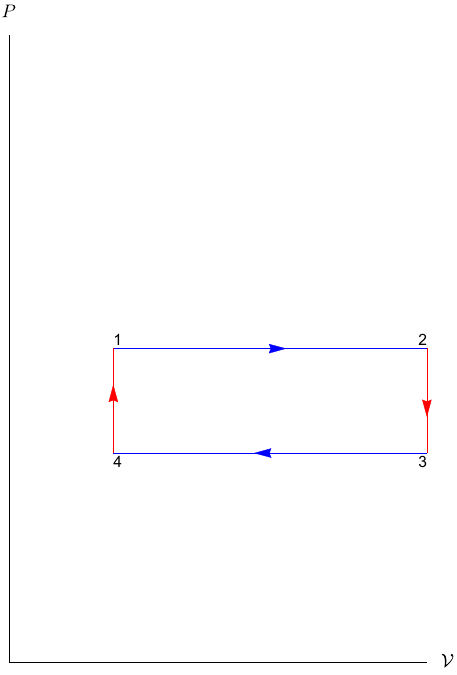}
		\caption{Rectangular cycle}
	\end{subfigure}
	\caption{$ P-{\cal V} $ diagram of thermodynamic cycles for the heat engine.}
	\label{fig5}
\end{figure}

\begin{figure}
	\centering
	\begin{subfigure}[b]{.496\linewidth}
		\includegraphics[width=\linewidth]{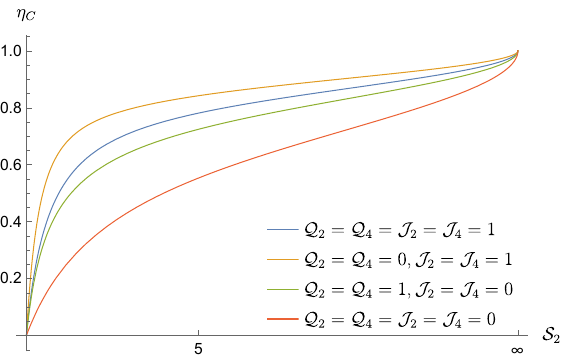}
		\setcounter{subfigure}{0}%
		\caption{}
	\end{subfigure}
	\begin{subfigure}[b]{.496\linewidth}
		\includegraphics[width=\linewidth]{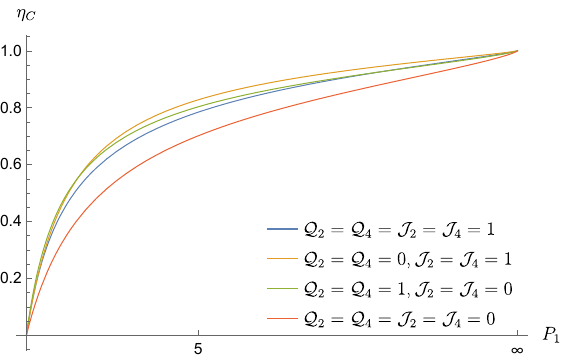}
		\caption{}
	\end{subfigure}
	\caption{Carnot heat engine efficiency as a function of (a) the entropy $ \mathcal{S}_2 $ and (b) pressure $ P_1 $, at $ {\cal S}_1=P_4=P_1={\cal S}_2=1 $.}
	\label{fig6}
\end{figure}

Among the various classical thermodynamic cycles, the Carnot cycle (see left panel of Fig. \ref{fig5}) stands out as the simplest and most idealized example. This cycle consists of two isothermal processes, one at the higher temperature $ T_H $, during which the system absorbs heat through isothermal expansion, and the other at the lower temperature $ T_C $, during which heat is released via isothermal compression. These isothermal segments are connected by two adiabatic processes (no heat exchange). The efficiency of a Carnot engine is given by:
\bea
\eta_{C}&=&1-\frac{T_C}{T_H}
\nn\\&=&1-\frac{T({\cal S}_1, P_4, {\cal J}_4, {\cal Q}_4)}{T({\cal S}_2, P_1, {\cal J}_2, {\cal Q}_2)}
\eea
Substituting the Hawking tempereture from Eq. (\ref{temp}) into the above formula yields
\bea
\eta_{C}&\!\!\!\!\!=\!\!\!\!\!&1\nn\\&&-\left\{\sqrt[4]{P_1} \mathcal{S}_2^{3/2} \varUpsilon _{2 C} \left(\pi  \mathcal{Q}_4^2-8 P_4 \mathcal{S}_1^2\right) \left(8 P_4 \mathcal{S}_1^2+3 \pi  \mathcal{Q}_4^2\right){}^3 \left(\varUpsilon _{2 C}+32 \pi ^{3/2} \sqrt{2} \mathcal{J}_2^2 P_1^{3/2} \mathcal{S}_2\right){}^{5/2} \right.\nn\\&&\left[32 \pi ^{3/2} \sqrt{2} \mathcal{J}_4^2 P_4^{3/2} \mathcal{S}_1 \varUpsilon _{4 C}+32 \pi ^3 P_4 \mathcal{S}_1^2 \left(64 \mathcal{J}_4^4 P_4^2+729 \mathcal{Q}_4^6\right)+165888 \pi  P_4^3 \mathcal{Q}_4^2 \mathcal{S}_1^6\right.\nn\\&&\left.\left.+93312 \pi ^2 P_4^2 \mathcal{Q}_4^4 \mathcal{S}_1^4+110592 P_4^4 \mathcal{S}_1^8+2187 \pi ^4 \mathcal{Q}_4^8\right]\right\}\left\{\sqrt[4]{P_4} \mathcal{S}_1^{3/2} \varUpsilon _{4 C} \left(\pi  \mathcal{Q}_2^2-8 P_1 \mathcal{S}_2^2\right) \right.\nn\\&&\left(8 P_1 \mathcal{S}_2^2+3 \pi  \mathcal{Q}_2^2\right){}^3 \left(\varUpsilon _{4 C}+32 \pi ^{3/2} \sqrt{2} \mathcal{J}_4^2 P_4^{3/2} \mathcal{S}_1\right){}^{5/2} \left[32 \pi ^{3/2} \sqrt{2} \mathcal{J}_2^2 P_1^{3/2} \mathcal{S}_2 \varUpsilon _{2 C}\right.\nn\\&&+32 \pi ^3 P_1 \mathcal{S}_2^2 \left(64 \mathcal{J}_2^4 P_1^2+729 \mathcal{Q}_2^6\right)+165888 \pi  P_1^3 \mathcal{Q}_2^2 \mathcal{S}_2^6+93312 \pi ^2 P_1^2 \mathcal{Q}_2^4 \mathcal{S}_2^4\nn\\&&\left.\left.+110592 P_1^4 \mathcal{S}_2^8+2187 \pi ^4 \mathcal{Q}_2^8\right]\right\}^{-1},
\eea
where $ \varUpsilon _{2 C}=\varUpsilon({\cal S}_2, P_1, {\cal J}_2, {\cal Q}_2) $, and $ \varUpsilon _{4 C}=\varUpsilon({\cal S}_1, P_4, {\cal J}_4, {\cal Q}_4) $. This represents the maximum possible efficiency attainable by any heat engine; achieving a higher efficiency would constitute a violation of the second law of thermodynamics. The Carnot heat engine efficiency is plotted against the entropy $ \mathcal{S}_2 $ and pressure $ P_1 $, in Fig. \ref{fig6}. From this figure, we found that increasing the entropy and pressure makes the increasing of the Carnot heat engine efficiency. In the limit of that the entropy $ \mathcal{S}_2 $ and pressure $ P_1 $ go to the infinity, the efficiency should approach to $ 1 $.
\begin{figure}
	\centering
	\begin{subfigure}[b]{.496\linewidth}
		\includegraphics[width=\linewidth]{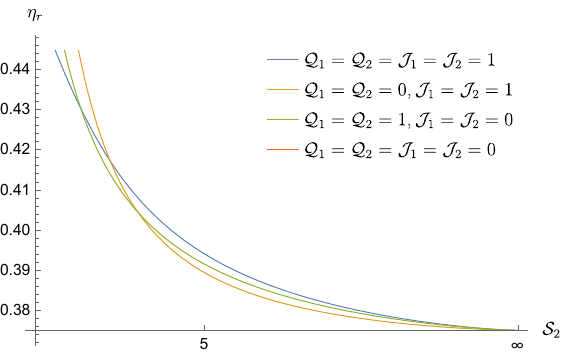}
		\setcounter{subfigure}{0}%
		\caption{}
	\end{subfigure}
	\begin{subfigure}[b]{.496\linewidth}
		\includegraphics[width=\linewidth]{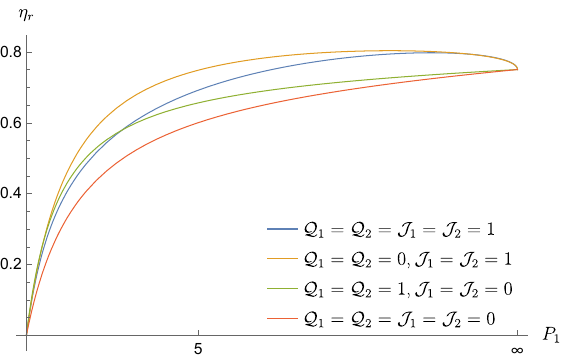}
		\caption{}
	\end{subfigure}
	\caption{Rectangular heat engine efficiency as a function of (a) the entropy $ \mathcal{S}_2 $ and (b) pressure $ P_1 $, at $ {\cal S}_1=1 $, $ P_1=2 $ and $ {\cal S}_2=1.2 $.}
	\label{fig7}
\end{figure}
\begin{figure}
	\centering
	\begin{subfigure}[b]{.496\linewidth}
		\includegraphics[width=\linewidth]{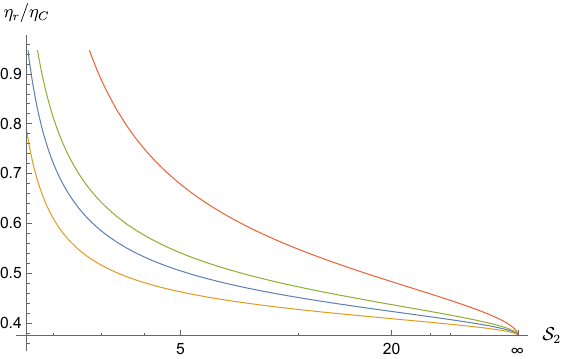}
		\setcounter{subfigure}{0}%
		\caption{}
	\end{subfigure}
	\begin{subfigure}[b]{.496\linewidth}
		\includegraphics[width=\linewidth]{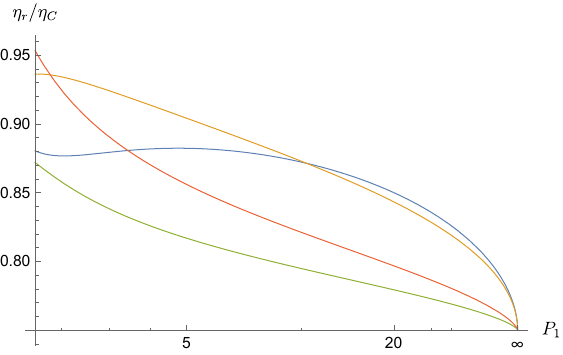}
		\caption{}
	\end{subfigure}
	\caption{The ratio $ \eta_r/\eta_C $ as a function of (a) the entropy $ \mathcal{S}_2 $ and (b) pressure $ P_1 $, at previous values of thermodynamic variables.}
	\label{fig8}
\end{figure}

One also can consider a rectangular engine cycle in $ P-{\cal V} $ plane (see right panel of Fig. \ref{fig5}) consisting of two isochores and two isobars path. The work done along the cycle reads
\bea
{\cal W} &=&
\oint Pd{\cal V} \nn\\&=& P_1 ({\cal V}_2 - {\cal V}_1) + P_4 ({\cal V}_4 - {\cal V}_3)\nn\\&=&(P_1-P_4)({\cal V}_2-{\cal V}_1).
\eea
The heat input along the upper isobar path can be calculated as
\bea\label{inheat}
Q_H = {\cal M}_2 - {\cal M}_1.
\eea
By substituting the black string mass from Eq. (\ref{mass}) into Eq. (\ref{inheat}), and applying the first line of Eq. (\ref{effi}), one finds
\bea
\eta_{r}&\!\!\!\!\!=\!\!\!\!\!& \left(P_1-P_4\right) \left(\left\{\left(32 \pi ^{3/2} \sqrt{2} \mathcal{J}_2^2 \mathcal{S}_2 P_1^{3/2}+\varUpsilon _{2 r}\right)^{1/2} \left[256 \pi ^{5/2} \sqrt{2} \mathcal{J}_2^2 \mathcal{Q}_2^2 \mathcal{S}_2 \varUpsilon _{2 r} P_1^{3/2}\right.\right.\right.
\nn\\&&+884736 \mathcal{S}_2^{10} P_1^5+1216512 \pi  \mathcal{Q}_2^2 \mathcal{S}_2^8 P_1^4+580608 \pi ^2 \mathcal{Q}_2^4 \mathcal{S}_2^6 P_1^3+93312 \pi ^3 \mathcal{Q}_2^6 \mathcal{S}_2^4 P_1^2\nn\\&&\left.\left.-8 \pi ^4 \mathcal{Q}_2^2 \left(729 \mathcal{Q}_2^6+2048 P_1^2 \mathcal{J}_2^4\right) \mathcal{S}_2^2 P_1-2187 \pi ^5 \mathcal{Q}_2^{10}\right]\right\}\nn\\&&\ls 108\ 2^{3/4} \sqrt{3} \sqrt[4]{\pi } P_1^{5/4} \sqrt{\mathcal{S}_2} \left(3 \pi  \mathcal{Q}_2^2+8 P_1 \mathcal{S}_2^2\right){}^5\rs^{-1}-
\nn\\&&\left\{\left(32 \pi ^{3/2} \sqrt{2} \mathcal{J}_1^2 \mathcal{S}_1 P_1^{3/2}+\varUpsilon _{1 r}\right)^{1/2} \left[256 \pi ^{5/2} \sqrt{2} \mathcal{J}_1^2 \mathcal{Q}_1^2 \mathcal{S}_1 \varUpsilon _{1 r} P_1^{3/2}+884736 \mathcal{S}_1^{10} P_1^5\right.\right.\nn\\&&+1216512 \pi  \mathcal{Q}_1^2 \mathcal{S}_1^8 P_1^4+580608 \pi ^2 \mathcal{Q}_1^4 \mathcal{S}_1^6 P_1^3+93312 \pi ^3 \mathcal{Q}_1^6 \mathcal{S}_1^4 P_1^2\nn\\&&\left.\left.-8 \pi ^4 \mathcal{Q}_1^2 \left(729 \mathcal{Q}_1^6+2048 P_1^2 \mathcal{J}_1^4\right) \mathcal{S}_1^2 P_1-2187 \pi ^5 \mathcal{Q}_1^{10}\right]\right\}\nn\\&&\left.\ls 108\ 2^{3/4} \sqrt{3} \sqrt[4]{\pi } P_1^{5/4} \sqrt{\mathcal{S}_1} \left(3 \pi  \mathcal{Q}_1^2+8 P_1 \mathcal{S}_1^2\right){}^5\rs^{-1}\right)
\nn\\&&\left\{\ls32 \pi ^{3/2} \sqrt{2} \mathcal{J}_2^2 \mathcal{S}_2 \varUpsilon _{2 r} P_1^{3/2}+36864 \mathcal{S}_2^8 P_1^4+55296 \pi  \mathcal{Q}_2^2 \mathcal{S}_2^6 P_1^3+31104 \pi ^2 \mathcal{Q}_2^4 \mathcal{S}_2^4 P_1^2\right.\right.\nn\\&&\left.+32 \pi ^3 \left(243 \mathcal{Q}_2^6+64 P_1^2 \mathcal{J}_2^4\right) \mathcal{S}_2^2 P_1+729 \pi ^4 \mathcal{Q}_2^8\rs\nn\\&&\ls2^{3/4} \sqrt{3} \sqrt[4]{\pi } \sqrt[4]{P_1} \sqrt{\mathcal{S}_2} \left(32 \pi ^{3/2} \sqrt{2} \mathcal{J}_2^2 \mathcal{S}_2 P_1^{3/2}+\varUpsilon _{2 r}\right){}^{3/2}\rs^{-1}\nn\\&&-\ls32 \pi ^{3/2} \sqrt{2} \mathcal{J}_1^2 \mathcal{S}_1 \varUpsilon _{1 r} P_1^{3/2}+36864 \mathcal{S}_1^8 P_1^4+55296 \pi  \mathcal{Q}_1^2 \mathcal{S}_1^6 P_1^3+31104 \pi ^2 \mathcal{Q}_1^4 \mathcal{S}_1^4 P_1^2\right.\nn\\&&\left.+32 \pi ^3 \left(243 \mathcal{Q}_1^6+64 P_1^2 \mathcal{J}_1^4\right) \mathcal{S}_1^2 P_1+729 \pi ^4 \mathcal{Q}_1^8\rs\nn\\&&\left.\ls2^{3/4} \sqrt{3} \sqrt[4]{\pi } \sqrt[4]{P_1} \sqrt{\mathcal{S}_1} \left(32 \pi ^{3/2} \sqrt{2} \mathcal{J}_1^2 \mathcal{S}_1 P_1^{3/2}+\varUpsilon _{1 r}\right){}^{3/2}\rs^{-1}\right\}^{-1},
\eea
where $ \varUpsilon _{1r}=\varUpsilon({\cal S}_1, P_1, {\cal J}_1, {\cal Q}_1) $, and $ \varUpsilon _{2r}=\varUpsilon({\cal S}_2, P_1, {\cal J}_2, {\cal Q}_2) $. The heat engine efficiency is plotted against the entropy $ \mathcal{S}_2 $ and pressure $ P_1 $, in Fig. \ref{fig7}. From this figure, we found that increasing the entropy makes the decreasing of the rectangular heat engine efficiency to the asymptotic value $ 0.375 $ of the uncharged and non-rotating case, while increasing the pressure makes the increasing of the rectangular heat engine efficiency. 

To gain an intuitive understanding, in Fig. \ref{fig8}, the ratio $ \eta_r/\eta_C $ is plotted versus the entropy $ {\cal S}_2 $ and pressure $ P_1 $. It can be seen that the ratio between rectangular and Carnot efficiency will monotonically decrease with the growth of $ {\cal S}_2 $ and $ P_1 $.

\section{Conclusion}\label{conc}

In recent years, black hole thermodynamics within the extended phase space has garnered significant attention. In this framework, the cosmological constant in AdS spacetime is treated as a dynamical quantity, corresponding to a variable thermodynamic pressure. This reinterpretation leads to equations of state for black holes that resemble those of non-ideal fluids, revealing a wealth of thermodynamic phenomena. 

One such process is the throttling (or Joule-Thomson) effect, a fundamentally irreversible process in which the enthalpy of a fluid remains constant between initial and final states. When applied to black holes in the extended phase space, this process is understood as both adiabatic and isenthalpic, where the black hole mass (interpreted as enthalpy) remains invariant.

In this work, we conduct a systematic investigation of the throttling process for asymptotically AdS charged and rotating black strings, a class of charged, rotating and cylindrically symmetric black holes in four-dimensional spacetime. While earlier studies \cite{Okcu:2016tgt,Okcu:2017qgo,Zhao:2018kpz} have primarily focused on black hole solutions with spherical horizon topology, we identify that the primary limitation in extending such analyses to black holes with non-spherical horizon topology.

We analyze the JT coefficient, inversion temperature, inversion and isenthalpic curves of black strings, analytically. Our findings indicate that, there are no maximum inversion temperatures-only minimum inversion temperatures exist, which is zero.

In this study, we also treat the black strings as a heat engine. The thermodynamic cycle considered here, illustrated on a $ P–{\cal V} $ diagram, consists of two isobaric and two isochoric processes. Our results indicate that the efficiency may either increase or decrease as the entropy $ {\cal S}_2 $ or Pressure $ P_1 $ grows.

We aim to offer a comprehensive picture of the throttling process and heat engine efficeincy for black strings and propose that our methods can be readily extended to more complex black holes, in particular, with non-spherical horizon topology.

It is worth noting that some significant aspect remains unexplored in this paper. For example, the impact of non-liniar parameter on the throttling process and heat engine efficeincy of blac strings, when non-liniar electrodynamics theories are included instead of Maxwell theory. We reserve this intricate and promising topic for future work.



\providecommand{\href}[2]{#2}\begingroup\raggedright
\endgroup
\end{document}